\documentclass[sigconf]{acmart}
\AtBeginDocument{%
  }

\renewcommand\footnotetextcopyrightpermission[1]{}
\setcopyright{none}
\acmConference{}{}{}
\acmBooktitle{}
\acmPrice{}
\acmDOI{}
\acmISBN{}
\begin{document}

\title{SWE-Shepherd: Advancing PRMs for Reinforcing Code Agents}

\author{Mahir Labib Dihan}
\affiliation{%
  \institution{Bangladesh University of Engineering and Technology (BUET)}
  \country{Bangladesh}
}
\email{mahirlabibdihan@gmail.com}

\author{Md Ashrafur Rahman Khan}
\affiliation{%
  \institution{Bangladesh University of Engineering and Technology (BUET)}
  \country{Bangladesh}
}
\email{ashrafurkhan37@gmail.com}

\renewcommand{\shortauthors}{Dihan et al.}

\begin{abstract}
Automating real-world software engineering tasks remains challenging for large language model (LLM)-based agents due to the need for long-horizon reasoning over large, evolving codebases and making consistent decisions across interdependent actions. Existing approaches typically rely on static prompting strategies or handcrafted heuristics to select actions such as code editing, file navigation, and test execution, but they lack fine-grained feedback on intermediate decisions. This leads to inefficient exploration, error propagation, and brittle solution trajectories. To address this limitation, we propose \textbf{SWE-Shepherd}, a framework that introduces Process Reward Models (PRMs) to provide dense, step-level supervision for repository-level code agents. Using trajectories from \textbf{SWE-Bench}, we construct an action-level reward dataset and train a lightweight reward model on a base LLM to estimate the usefulness of intermediate actions. During inference, the PRM evaluates candidate actions and guides the agent toward higher-reward decisions without requiring full reinforcement learning. Experiments on \textbf{SWE-Bench Verified} demonstrate improved interaction efficiency and action quality, while also highlighting challenges in aligning intermediate rewards with final task success.
\end{abstract}

\maketitle

\section{Introduction}

Automating real-world software engineering tasks remains a major challenge for large language models (LLMs). Tasks such as bug fixing, code modification, and test-driven development require long-horizon reasoning, interaction with large and evolving codebases, and consistent decision-making across sequences of interdependent actions. Existing LLM-based agents typically rely on handcrafted heuristics or static prompting strategies to select actions such as reading files, editing code, or executing tests. While effective in constrained settings, these approaches often lack mechanisms to evaluate intermediate decisions, leading to inefficient exploration, error propagation, and brittle solutions.

To address these limitations, we introduce \textbf{SWE-Shepherd}, a framework that operationalizes \textbf{Process Reward Models (PRMs)} for repository-level code agents. Instead of relying solely on sparse signals of final task success, SWE-Shepherd converts agent trajectories into dense, step-level supervision by assigning scalar rewards to intermediate actions according to their estimated contribution toward resolving the issue. Using trajectories collected from \textbf{SWE-Bench}, we construct a dataset of action-level reward annotations and train a lightweight reward model on top of a base LLM.

At inference time, the PRM evaluates multiple candidate actions and guides the agent toward those predicted to be more useful, enabling reward-guided search without requiring full reinforcement learning. This design positions PRMs as a practical middle ground between supervised imitation learning and RL-based optimization: they provide dense behavioral feedback while remaining simple to train and deploy.

Our goal is not only to improve task performance, but also to study whether process-level supervision can produce more efficient and interpretable decision-making in code agents. Through experiments on SWE-Bench Verified, we show that PRM guidance reduces interaction steps and alters agent behavior, while also revealing important alignment challenges between intermediate rewards and final task success.

\noindent
\textbf{Codebase:} \url{https://github.com/mahirlabibdihan/swe-shepherd} \\
\textbf{Dataset:} \url{https://huggingface.co/datasets/mahirlabibdihan/swe-prm-collection} \\
\textbf{Model:} \url{https://huggingface.co/mahirlabibdihan/swe-shepherd-1.5b}

\begin{figure}
    \centering
    \includegraphics[width=1\linewidth]{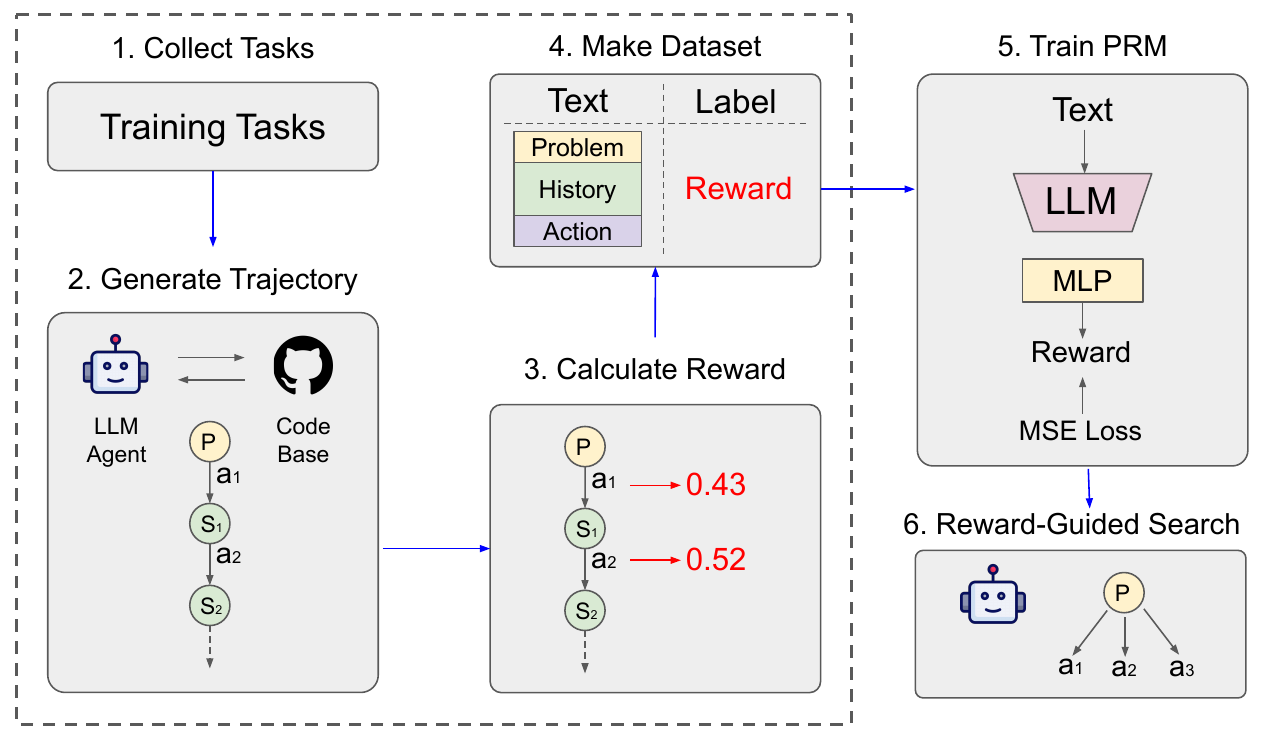}
    \caption{Overview of the SWE-Shepherd framework. The framework consists of six stages. (1) Training tasks are collected from SWE-Bench. (2) An LLM-based agent tries to solve the tasks by interacting with the codebase and generate solution trajectories composed of intermediate reasoning steps and actions. (3) Each intermediate step is assigned a scalar reward reflecting its contribution toward solving the task. (4) A dataset is constructed where the input text (problem description, execution history, and current action) is paired with the computed reward labels. (5) A Process Reward Model (PRM) is trained by feeding text representations from the LLM into an MLP head and optimizing with mean squared error (MSE) loss to predict step-level rewards. (6) During inference, the trained PRM guides reward-aware search, prioritizing higher-quality intermediate steps to improve problem-solving performance.}
    \label{fig:overview}
\end{figure}

\section{Related Work}




Recent work has explored the use of LLMs for autonomous software engineering, with benchmarks such as \textbf{SWE-Bench}\citep{jimenezswe} and SWE-Bench Verified\citep{openai_swebench_verified_2024} highlighting the difficulty of long-horizon reasoning, repository-scale context understanding, and test-driven correctness.

Several approaches attempt to improve agent behavior using \emph{reinforcement learning (RL)}. Methods such as SWE-RL\citep{weiswe} and Agent-RLVR\citep{da2025agent} apply policy optimization or environment-driven reward signals to address sparse supervision and improve trajectory planning. In contrast, other work focuses on \emph{data synthesis and supervised learning}, for example SWE-Synth\citep{pham2025swe}, which generates structured bug-fix trajectories to provide dense training signals without RL.

Process-level reward modeling has recently emerged as an alternative paradigm for guiding multi-step agents. Web-Shepherd\citep{chae2025webshepherd}, for instance, predicts the utility of intermediate steps to steer decision-making in web navigation tasks. Our work extends this idea to repository-level software engineering by constructing PRMs tailored to code-editing environments and studying their effectiveness as a lightweight alternative to RL-based training.

\section{Methodology}

This section presents the SWE-Shepherd pipeline, covering task collection, trajectory generation, reward computation, dataset construction, process reward model training, and reward-guided inference (Figure \ref{fig:overview}).

\noindent\textbf{Task Collection}
We use the SWE-Bench dataset, which contains real-world GitHub issues, repository snapshots, and test suites. Each task includes a repository, base commit, problem statement, and tests. Out of 2,294 tasks, 500 form the SWE-Bench Verified subset used for evaluation, while the remaining 1,794 tasks are used for training.

\noindent\textbf{Trajectory Collection}
A baseline LLM-based agent generates trajectories for each task, consisting of alternating actions (reading files, editing code, running tests) and observations (outputs, file contents, test results). These trajectories capture the agent’s reasoning and decision-making.

\noindent\textbf{Reward Computation.}
We assign intermediate rewards based on heuristics reflecting progress toward task resolution, including successful execution, relevant file access, target file modification, test results, and avoidance of repetitive actions. Discounted cumulative rewards capture long-term contributions.

\noindent\textbf{Dataset Construction.}
Trajectories and rewards are converted into supervised training samples: problem statement, recent interaction history, candidate action, and scalar reward (normalized to $[0,1]$). The dataset contains over 15,000 samples covering diverse tasks.

\noindent\textbf{Process Reward Model Training.}
The Process Reward Model (PRM) predicts the reward of candidate actions given context. Built on a pretrained language model, the final hidden token is projected to a scalar. We use qLoRA for parameter-efficient fine-tuning and train with mean squared error (MSE) loss.

\noindent\textbf{Reward-Guided Inference.} 
During inference, the PRM scores candidate actions at each step, and the agent selects the highest-scoring action. This reward-guided selection improves decision-making without requiring reinforcement learning, continuing until the issue is resolved or a step limit is reached.

\section{Experiments}

\subsection{Experimental Setup}

We evaluate SWE-Shepherd on 100 tasks sampled from SWE-Bench Verified. Each task requires generating a patch that resolves the issue and passes all associated tests. The agent is limited to a maximum of 30 interaction steps.


\noindent We compare:
\begin{itemize}
\item \textbf{mini-SWE-Agent \citep{miniSWEAgentDocs2026}:} A strong LLM agent that follows a sequential decision process without explicit search or reward modeling.

\item \textbf{SWE-Search \citep{antoniadesswe}:} A search-based framework that augments software agents with Monte Carlo Tree Search (MCTS) to explore multiple candidate action trajectories before committing to a solution.

\item \textbf{SWE-Shepherd (Ours):} mini-SWE-Agent augmented with the trained Process Reward Model (PRM) for action scoring and selection.
\end{itemize}



\subsection{Results}

\begin{table}[h]
\centering
\small
\begin{tabular}{lccc}
\toprule
\textbf{Method} & \textbf{\% Resolved} & \textbf{Avg. \$} & \textbf{Avg. Steps} \\
\midrule
SWE-Search & 31\% & 0.274 & -- \\
mini-SWE-Agent & 57\% & 0.029 & 15.2 \\
SWE-Shepherd (Ours) & 51\% & 0.053 & 12.2 \\
\bottomrule
\end{tabular}
\caption{Performance on SWE-Bench Verified using \texttt{gpt-5-mini} (100 tasks, max 30 steps).}
\end{table}

Compared to SWE-Search, both mini-SWE-Agent and SWE-Shepherd achieve substantially higher resolution rates while requiring significantly lower cost, highlighting the effectiveness of iterative agent-based interaction over expensive search-based exploration. SWE-Shepherd reduces the number of interaction steps, indicating more directed exploration. However, it yields a modest drop in resolution rate, suggesting that locally high-reward actions do not always translate to globally correct patches.

\subsection{Reward Analysis}

\begin{table}[h]
\centering
\small
\begin{tabular}{lc}
\toprule
\textbf{Task Outcome} & \textbf{Average Reward} \\
\midrule
Resolved Tasks & 0.4894 \\
Unresolved Tasks & 0.4818 \\
\bottomrule
\end{tabular}
\caption{Average reward for resolved vs. unresolved tasks.}
\end{table}

The small difference in rewards indicates that the current reward function only weakly correlates with task success.

\subsection{Discussion}

These findings highlight a key challenge in process-level supervision: intermediate behavioral signals are easier to model than repository-level correctness. While PRMs encourage efficient trajectories, misalignment between heuristic rewards and final task success can bias agents toward locally consistent but incomplete solutions.

\section{Future Work}

Future work includes:
\begin{itemize}
\item Improving reward modeling to better correlate intermediate rewards with task success.
\item Expanding the training dataset with more diverse and complex tasks.
\item Exploring hybrid approaches that combine reward-guided search with reinforcement learning to further improve resolution rates.
\item Investigating adaptive step limits and candidate action generation strategies to enhance efficiency and success.
\end{itemize}

\bibliographystyle{ACM-Reference-Format}
\bibliography{sample-base}

\end{document}